\def\al{{$i$-Al-Pd-Mn}}
\def\ef{{$E_F$}}
\documentclass[amsmath,twocolumn,amssymb,aps,prb]{revtex4-2}
\usepackage{parskip}
\usepackage{graphicx}
\usepackage{hyperref}
\usepackage{hypernat}
\usepackage{natbib}
\hypersetup{pdfencoding=auto,colorlinks=true,linkcolor=blue,citecolor=blue}
\usepackage{appendix}
\usepackage{etoolbox}
\usepackage{dcolumn}
\usepackage{bm}
\usepackage[mathlines]{lineno}
\usepackage[depth=subsection]{bookmark}
\usepackage{comment}
\usepackage{xr} 
\newcommand{\srb} {\textcolor{black}}

\pdfoutput=1 

\begin{document}
	
	\title{Quasiperiodic gallium adlayer on \textit{i}-Al-Pd-Mn}

	\author{Pramod Bhakuni$^1$, 	Marian Kraj\v{c}\'i$^{2,*}$,  Sudipta Roy Barman$^{1,\dagger}$}
	\affiliation{$^{1}$UGC-DAE Consortium for Scientific Research, Khandwa Road,  Indore - 452001, Madhya Pradesh, India}
	\affiliation{$^{2}$Institute of Physics, Slovak Academy of Sciences, D\'ubravsk\'a  cesta 9, SK-84511 Bratislava, Slovak Republic}
	
	
	\begin{abstract}
Using scanning tunneling microscopy (STM), low energy electron diffraction (LEED), and density functional theory (DFT), we demonstrate the formation of quasicrystalline gallium adlayer on icosahedral ($i$)-Al-Pd-Mn.  Quasiperiodic motifs are evident in the STM topography images, including the Ga white flower (GaWF) and $\tau$ inflated GaWF ($\tau$-GaWF), where $\tau$ is the golden mean. A larger and more complicated  ring motif is also identified, comprised of a bright center and an outer ring of pentagons. \srb{The fast Fourier transform  of the STM images exhibits distinct quasiperiodic spots, thereby establishing quasiperiodicity on a length scale of  $\sim$350 nm.} Based on our DFT calculations, the preferred adsorption sites of Ga on \al\  are determined to be either the Mn atoms at the center of the Penrose P1 tile or the vertices of the  P1 tile containing Pd atoms at the center of a cluster of 5 Al atoms (5-Al). The GaWF motif is modeled by an inner 6 atom Ga cluster (6-Ga) around the central Mn atom and an outer ring of 5 Ga atoms adsorbed at the centers of the  5-Al clusters, both having pentagonal symmetry. The $\tau$-GaWF  motif is  modeled by the 6-Ga arranged on the $\tau$-P1 tiling, \srb{while the ring motif is modeled by Ga atoms adsorbed  at the center of 5-Al clusters above a Pd atom. 
~The  side lengths and diameters of the GaWF, $\tau$-GaWF, and the ring motifs are $\tau$ scaled and  show excellent agreement with 
~the DFT-based  models. 
~An additional indication of the quasiperiodic characteristics of the Ga monolayer is the 5-fold LEED patterns that were detected throughout the entire range of beam energy that was measured.} 

	\end{abstract}
	\maketitle

\section{Introduction}
\label{sec:intro}
Quasicrystals are a distinct type of material that display aperiodic order with rotational symmetries, including 5-fold, 8-fold,  10-fold, and \srb{12-fold}, which are  forbidden in crystals with translational order~\cite{penrose1974role, Levine1984}. The quasicrystalline phase was first identified in a binary alloy of Al$_6$Mn with icosahedral symmetry~\cite{Shechtman1984}, and since then it has been observed in a variety of systems, which include ternary and binary intermetallic compounds~\cite{Tsai2000}, nanoparticle superlattices~\cite{Talapin2009}, colloidal systems~\cite{Fischer2011}, \srb{perovskite barium titanate on Pt(111)}~\cite{Forster2013}, molecular assemblies~\cite{Xiao2012, Ye2016}, twisted bilayer graphene~\cite{Yao2018, Ahn2018, Maniraj2020}, chalcogenides~\cite{Cain2020}, elemental films~\cite{Franke2002,Sharma2013,Singh2020,Singh2023}, and even in naturally occurring minerals~\cite{Bindi2012, Bindi2022}. Inorganic quasicrystals exhibit unusual properties such as low thermal and electrical conductivity, low specific heat,  low frictional coefficient, and large hardness~\cite{Poon1992, Kang1993, Dubois1994, Park2005, Dubois2012}. Their stability has been related to the pseudogap i.e., a reduced density of states (DOS) around the Fermi level~\cite{Hafner1992, Nayak2012, Nayak2015,Sarkar2021prr}.  The possibility of higher order topological states occurring in quasiperiodic systems has been predicted by theory~\cite{Chen2020, Fan2021, Wang2022}. Our recent research utilizing hard x-ray photoemission spectroscopy and density functional theory (DFT) has established the occurrence of Anderson localization in icosahedral ($i$)-Al-Pd-Re~\cite{Sarkar2021}.

\begin{figure*}[!t] 
	\includegraphics[width=1.13\linewidth,keepaspectratio,trim={2cm 2cm 0 1cm },clip]{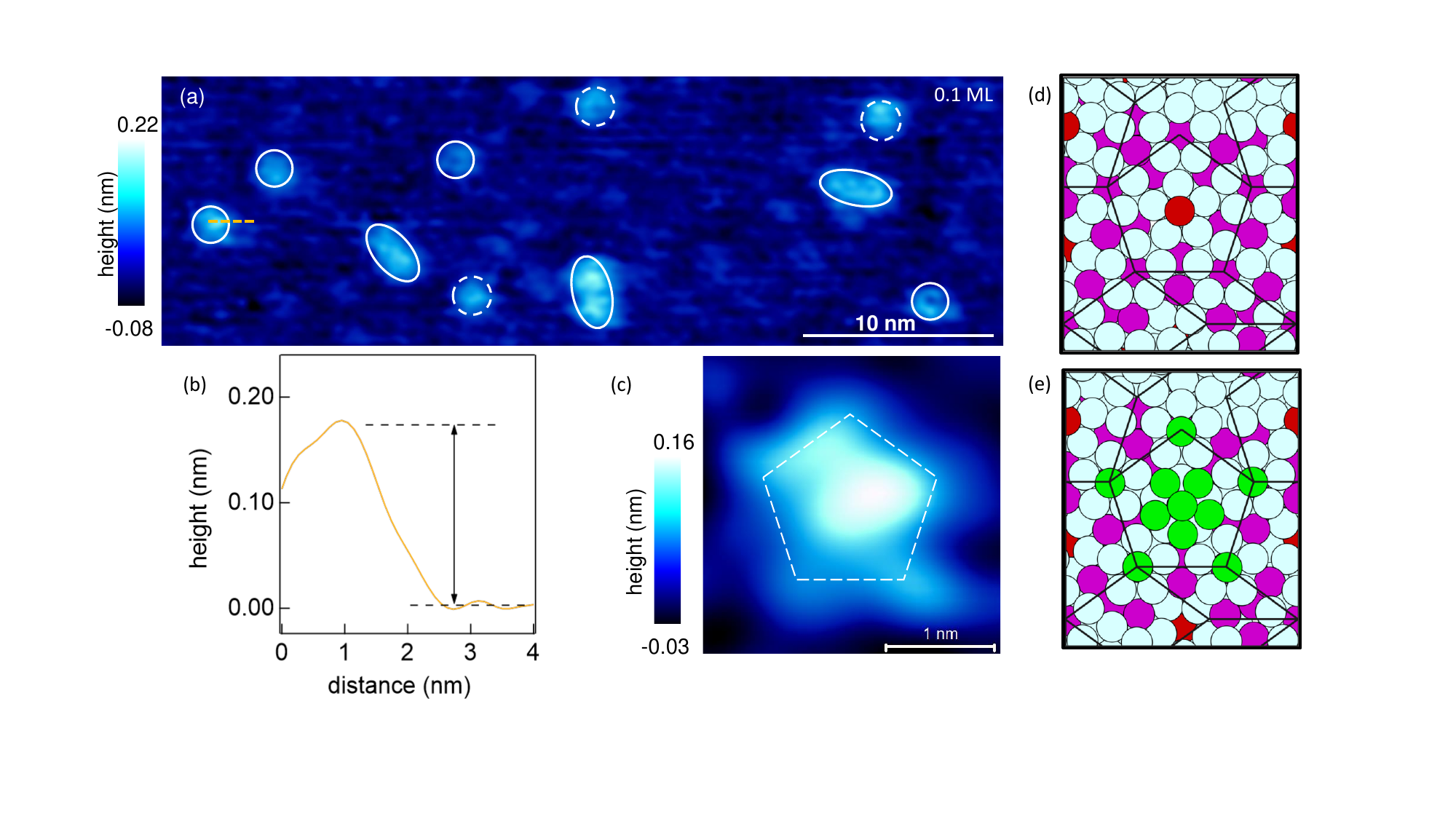}
	\caption{(a) STM topography image of 0.1 ML Ga/\al\, ($I_T$ = 0.9 nA , $U_T$ = -0.7 V) showing the Ga white flower (GaWF) motifs encircled by white circles and ovals, dashed circles show the  incomplete motifs. The color scale representing the	height  is shown on the left, zero corresponds to the bearing height.	(b) A height profile across the GaWF along the  orange dashed  line in panel \textbf{a}. (c) Zoomed view of a GaWF motif with a  white dashed pentagon  overlaid on it.  (d) The atomic structure of the surface plane (2.3 nm$\times$2.4 nm) 
~of the 5-fold \al\, derived from the 2/1 approximant to the bulk. The white, magenta, and red colored circles represent the Al, Pd, and Mn atoms, respectively.  \srb{The Penrose P1 tiling is shown by black lines.} (e) An 11 Ga atom DFT  based model of the GaWF centered on the Mn site of the  5-fold \al. The green colored circles represent the Ga atoms.} 
	\label{nuclstm}
\end{figure*}

Since more than two decades, researchers have studied elemental quasicrystalline films to investigate the impact of aperiodic order on the physical and electronic properties that are independent of the chemical complexity of the ternary quasicrystals. But to date, only a few elements have demonstrated quasiperiodicity~\cite{Franke2002, Cai2003, Shimoda2004,Ledieu2005, Shukla2006, Smerdon2008, Ledieu2008, Shukla2009, Ledieu2009, Sharma2013,Singh2020,Singh2023}. Franke \textit{et al.} discovered that Bi and Sb are quasiperiodic on \al\, and decagonal Al-Ni-Co quasicrystalline substrates up to 1 monolayer (ML) coverage using  low energy electron diffraction (LEED) and He atom scattering~\cite{Franke2002}. Shukla \textit{et al.}~\cite{Shukla2009} showed  that  Na forms a  regular 5-fold quasiperiodic bilayer on \al.  Pb/$i$-Ag-In-Yb was reported to exhibit 5-fold growth isostructural with the \srb{$i$-Ag-In-Yb} substrate to a maximum height of 0.7 nm~\cite{Sharma2013}.   \srb{Recent research utilizing scanning tunneling microscopy (STM), LEED, and DFT has demonstrated that a 4 nm thick Sn adlayer deposited on the fivefold surface of an $i$-Al-Pd-Mn substrate exhibits long range quasiperiodic order as a metastable realization of an elemental, clathrate family quasicrystal~\cite{Singh2020}.  Long range decagonal clathrate quasiperiodic ordering of Sn thin films on $d$-Al-Ni-Co has also been observed up to 0.9 nm, with partial retention of decagonal structural correlations up to 10 nm film thickness~\cite{Singh2023}.}

Gallium is an interesting element due to its relatively low melting point (303 K) and its complex solid-state structure. The latter produces a notable dip at the Fermi level (\ef), which is believed to be the consequence of partial covalent bonding~\cite{Hafner1990, Barman1995}. Recently, gallanene i.e., a honeycomb monolayer of Ga that is analogous to graphene, has been reported~\cite{Kochat2018, Tao2018, Badalov2018, Wundrack2021, Kutana2022}. However, Ga deposition on single crystalline metal surfaces has been scarcely studied~\cite{Jeliazova2003},  and there has been  no study on any quasicrystalline substrates to date.  Ga  could be a possible candidate since its surface energy  (0.55 J/m$^2$)~\cite{Tran2016} is less than that reported for the \textit{i}-Al-Pd-Mn surface (0.82 J/m$^2$)~\cite{Dubois2006},  indicating the possibility of  layered  growth.  \srb{In this work, we use STM, LEED, and density functional theory (DFT) to reveal the occurrence of long range quasiperiodicity in a gallium monolayer deposited at room temperature (RT) on \al.  The quasiperiodic motifs such as Ga white flower (GaWF), $\tau$ scaled GaWF ($\tau$-GaWF), and the ring observed from STM  have been modeled by  our DFT calculations. 	The fast Fourier transform  of the large area STM images exhibits distinct quasiperiodic spots. }

\section{Methods} 
	\label{sec:methods}
	The STM measurements were carried out using a variable temperature STM  from Scientaomicron GmbH  at a base pressure of 5$\times$10$^{-11}$ mbar. All the STM measurements were performed at RT in the constant current mode using a tungsten tip that was prepared by sputtering and the voltage pulse method. The tip was biased, and the sample was kept at the ground potential. STM images were recorded for varying bias voltages. \srb{The Fast Fourier transform  (FFT) was averaged for 2-4 STM images, as in our previous work~\cite{Singh2023}.} The STM images are shown after low-pass Fourier transform filtering \srb{(Fig.~S1 of the Supplementary material (SM)~\cite{supple})}. The zero in the z-scale of the STM image corresponds to the most frequently occurring height (bearing height) based on the height histogram. In a height profile, the difference of the average $z$ corrugation on the Ga adlayer and the substrate provides the height of the former~\cite{Sadhukhan2019}. 
	~The average height for a particular deposition is determined by fitting a Gaussian curve to the height distribution obtained from more than 50 height profiles derived from various parts of the STM images.  LEED and AES were performed using a four-grid rear view optics and a hemispherical retarding field analyzer, respectively. The \textit{I-V} curves were extracted using the EasyLeed software~\cite{Mayer2012}. 
	
		The 5-fold surface of the monocrystalline \textit{i}-Al-Pd-Mn 
~		was polished using a 0.25 $\mu$m diamond paste 
~before inserting it into the ultra-high vacuum chamber. The polished \textit{i}-Al-Pd-Mn surface was treated \textit{in-situ} by repeated cycles of Ar$^+$ ion sputtering at 1-2 keV for 30-60 min and annealing to 970 K for 2-2.5 hr to produce an atomically clean surface with a composition similar to the bulk~\cite{Fourne2000,Maniraj2014}.  Gallium of 99.99$\%$ purity was evaporated using a water-cooled Knudsen cell~\cite{Shukla2004} operating at 1050 K  at a pressure  better than 2$\times$10$^{-10}$ mbar.  The substrate surface was freshly prepared for each deposition. 

The DFT calculations to probe the interaction of Ga atoms with the \al~ surface were performed using the method of the Vienna Ab initio Simulation Package (VASP)~\cite{Kresse1996, Kresse1996prb, Kresse1999}. The models of the surface have been derived from the Katz-Gratias-Boudard model~\cite{Kraj1995} of bulk \al. The atomic structure of the 5-fold surface is derived from the icosahedral approximants by cleaving at a plane perpendicular to one of the 5-fold axes. Details of construction of the surface models can be found in our previous papers~\cite{Kraj2006, Kraj2008, Ledieu2009, Kraj2010}.  
~\srb{The adsorption energy ${E_A[\rm Ga]}$ of  Ga adatom is defined with respect to the energy in the crystalline structure as 
$E_A[\rm Ga]$ = $E_{tot}[\rm Ga]$$-$$E_{surf}$+$N_A$$\times$$\mu[\rm Ga]$, 
where $E_{tot}$ is the energy of the \al\ surface model~\cite{Kraj1995,Ledieu2009} with the adsorbed adatoms.  $E_{surf}$ is the energy of the model \al\ surface representing the atomic structure of the clean quasicrystalline 5-fold surface. ${N_A}$ is the number of adatoms in the cluster on the surface. 
~$\mu[\rm Ga]$ is the chemical potential of Ga in the crystalline structure, $\mu$[Ga] = $-E_{cryst}$[Ga.oP8]=  2.934 eV. 
~The adsorption energy per atom $E_a$ is {$E_a$[Ga] = $E_A$[Ga]/$N_A$}.}

	\section{\label{results}Results and discussion}
\subsection{\label{subsec:submonolayer_stm}Nucleation of  Ga on \al}

  A STM topography image  for a deposition time ($t_d$) of 2 min  results in a coverage of 0.1 ML Ga on \al. The coverage is determined by the fraction of the total area covered by the Ga clusters that are identified by the bright regions  in Fig.~\ref{nuclstm}(a).  The clusters have pentagonal symmetry that resembles flowers with five petals and are highlighted by white circles and ovals.    We  refer to these  as Ga white flowers (GaWFs).  The height of the GaWFs is determined by subtracting the average $z$ corrugation of the  substrate  from that of the GaWF, as illustrated in Fig.~\ref{nuclstm}(b).  In Fig.~S2(a) 
  ~of the SM~\cite{supple}, a distribution of the heights obtained from several height profiles is fitted with a Gaussian function. The position of its maximum gives the average height of the GaWF to be 0.18$\pm$0.03 nm.     A zoomed view of a GaWF motif is shown  in Fig.~\ref{nuclstm}(c). The length of the sides of the  pentagon (white dashed  lines) joining its petals  has been determined for several  motifs. Its average length turns out to be 0.79$\pm$0.1 nm,   and the corresponding distribution is shown in Fig.~S2(b) 
  ~of SM~\cite{supple}.

To understand the origin of the GaWF, we have performed DFT calculation for the adsorption of Ga atoms on the  5-fold \al\, surface that  was modeled as the 2/1 approximant [Fig.~\ref{nuclstm}(d)].   The 5-fold \al\, surface is superposed with  the Penrose P1 tiling (black lines)~\cite{Kraj2006}. The  P1 tiling has an edge length ($a_0$) of  0.776 nm.  The  Mn atoms (red circles) appear at the centers of the pseudo-Mackay clusters (pMC). The vertices of the P1 pentagon are occupied by 5 Pd atoms (magenta circles). This figure shows that the surface Mn atoms are  in the center of the P1 tiling, around which the well known white flower motifs of the substrate form~\cite{Papadopolos2002, Barbier2002}.

Quasicrystalline surfaces offer a greater number of inequivalent adsorption sites compared to ordinary crystals, allowing for the formation of various adatom configurations.  On the 5-fold \al\, surface, there are two kinds of regular sites that preferably adsorb Ga atoms. The first site is the center of the surface 5-Al cluster that has  a Pd atom. These Pd atoms can be 0.048 nm or 0.126 nm  below the surface plane. \srb{Therefore, there are 2 kinds of the Pd sites and only one can adsorb the adatoms.} The Ga atoms are adsorbed preferably only at the centers of the 5-Al cluster with the  deeper  Pd atom. \srb{These deeper Pd atoms are the centers of the Bergman clusters~\cite{Kraj1995}}. In Fig.~\ref{nuclstm}(d), these sites are at the vertices of the P1 tiles and are also marked  by white dots in Fig.~\ref{motif}(f) and Fig. S3(b) of SM. The binding energy of a \srb{single} Ga atom in such sites  is   $-$1.357 eV  (see Table~\ref{table1}). The second preferable adsorption site for \srb{a Ga adatom}   is the central Mn atom at the surface.  The main reason for the higher reactivity of these Mn atoms is their  low coordination: In the bulk pMC cluster, in the first shell  around the Mn atom, there are only 7 nearest neighbors~\cite{Kraj1995, Kraj2005}. The binding energy of Ga atoms at the surface Mn atoms  is between $-$1.2 eV and $-$1.4 eV. 
	
In previous studies~\cite{Ledieu2009, Kraj2005, Kraj2010, Maniraj2014,Singh2020, Smerdon2008}, it was observed that the adatoms (Bi, Pb, and Sn) form WF clusters around the surface Mn atoms. \srb{Our DFT calculations show that Ga atoms adsorbed on the 5-fold \al\, surface also form clusters similar to the previously observed WFs, but the internal structure  is distinct because of the notably different electronic structure of Ga~\cite{Hafner1990, Barman1995}.} In the previous studies~\cite{Ledieu2009, Kraj2005, Kraj2010, Maniraj2014,Singh2020, Smerdon2008}, the adatoms form the WF cluster, in which the outer and inner rings are oriented in the same direction and are therefore interconnected to form a starfish-like configuration.  The GaWF cluster also consists of the outer ring of five Ga atoms, each adsorbed at the vertices of the P1 pentagon that are the centers of the reactive 5-Al clusters [Fig.~\ref{nuclstm}(e)] that is the same as in the previous WFs. 
However, in contrast, 
the central pentagon of Ga adatoms has an opposite orientation compared to the pentagon of the atoms in the outer ring. Moreover,  \srb{to stabilize this central	pentagon, there must be an additional Ga atom } at the center, making it a 6 atom Ga cluster (6-Ga), while retaining the pentagonal symmetry. \srb{The energies calculated for different Ga clusters are discussed later and compared in Table \ref{table1}.}  The opposite orientation of the pentagonal rings reflects the lower attractive interaction of adatoms between the outer ring and central 6-Ga cluster.  \srb{In the case of the Ga adatoms, the outer and inner 	rings can also exist  independently. } The formation of the central 6-Ga cluster is slightly (-0.16 eV/Ga atom) preferred  over the formation of both rings simultaneously.  The central Ga atom in the 6-Ga cluster is  elevated by 0.1 nm  compared to the  five Ga neighbors. Therefore, the center of the GaWF cluster may appear brighter in the STM images than the outer ring. Also, on the 6-Ga cluster at the center, additional Ga atoms can be adsorbed,  which can lead to a brighter center and an irregular shape. We also note that, due to stochastic reasons, some GaWF motifs have a dark center, which could be caused by the absence of the  6-Ga cluster  at the center. 

It should be noted   that the edge length of the P1 tiling $a_0$ (= 0.776 nm) is in excellent agreement with the side length of the GaWF from STM (0.79$\pm$0.1 nm). Additionally,  the typical distance between two neighboring Mn atoms  on which the 6-Ga clusters are  centered is  1.255 nm (= $\tau$$\times$$a_0$, where $\tau$= 1.618). This distance  also agrees well with the distance between the nearest neighbor GaWFs, as shown by the white ovals in Fig.~\ref{nuclstm}(a), where the  distance between their centers is $\sim$1.25 nm. This confirms that the 6-Ga clusters nucleate on the Mn atoms i.e., on the white flower motifs of  the substrate. 
	\vskip -0.3cm
		\begin{table}[!ht]
			\caption{\srb{	Adsorption energies per atom ${E_a}$ [eV] of Ga clusters 
				consisting from ${N_A}$ adatoms on the 5-fold \al\ surface.}}
			\bigskip
			\begin{tabular}{|c|c|c|c|c|}\hline
				~~~~~~~~& Single Ga & Outer 5-Ga & WF cluster & Inner 6-Ga  \\ \hline
				${N_A}$ & 1 & 5  & 11 & 6 \\ \hline
				${E_a}$ & $-$1.357 & $-$0.451  & $-$0.354 & $-$0.515 \\ \hline
			\end{tabular}
			\label{table1}
		\end{table}
\srb{A comparison of the adsorption energies of Ga adatoms on the 5-fold \al\  surface are presented in Table \ref{table1}. As mentioned above, the adsorption energy of $-$1.357 eV corresponds to a single  Ga adatom adsorbing at the reactive center of the 5-Al surface cluster above the Pd atom.
~The adsorption energy per atom ($E_a$) of the pentagonal 5-Ga  outer ring of adatoms adsorbed  in 	the five reactive 5-Al surface sites is $-$0.451 eV. The GaWF cluster with an $\rm{E_a}$ of $-$0.354 eV consists of the outer 5-Ga ring and the inner 6-Ga cluster.  
	~The $E_a$ of the central 6-Ga cluster formed alone around the surface Mn atom is the strongest ($-$0.515 eV). The adsorption of adatoms in the clusters is weaker compared to the adsorption of a single adatom because, in the cluster, in addition to bonding of adatoms to the surface, there are also interactions	between the adatoms. }
	
	\srb{ For comparison, in Discussion I of the SM, we also present the 	adsorption energies of the previously studied Sn white flower (SnWF) cluster~\cite{Singh2020} on the same 5-fold \al\ surface.  ${E_a}$ of the complete SnWF cluster is  $-$0.795 eV (see Table S1 of the SM), which is significantly larger than ${E_a}$ = $-$0.354 eV of the GaWF cluster.  This could be a possible reason for the somewhat worse quality of the GaWF motifs [Fig.~\ref{nuclstm}(a)] compared to the SnWF motifs on \al\,  [see Fig.~10(a) and Fig.~S8 of Ref.~\onlinecite{Singh2020}].  
	~Another possible reason  is that the STM in the present work has been performed at room temperature, which is close to the melting point of Ga, resulting in enhanced thermal effects.}

\srb{The 5-fold surface of \al\ is primarily composed of Al atoms, with the top two atomic planes of the 5/3-approximant consisting of 75.9\% Al, 19.6\% Pd, and 4.5\% Mn~\cite{Kraj2005}.  
~So, it can be interesting to compare the adsorption of Ga adatoms  on the crystalline Al(111) surface with the results in Table~I for the quasicrystalline surface. Discussion~II of SM~\cite{supple} shows that the reactivity of the adsorption sites on \al\  is significantly higher compared to Al(111).}

 	\begin{figure*}[!t] 
 	\includegraphics[width=1.3\linewidth,keepaspectratio,trim={6cm 1cm 0 0 },clip]{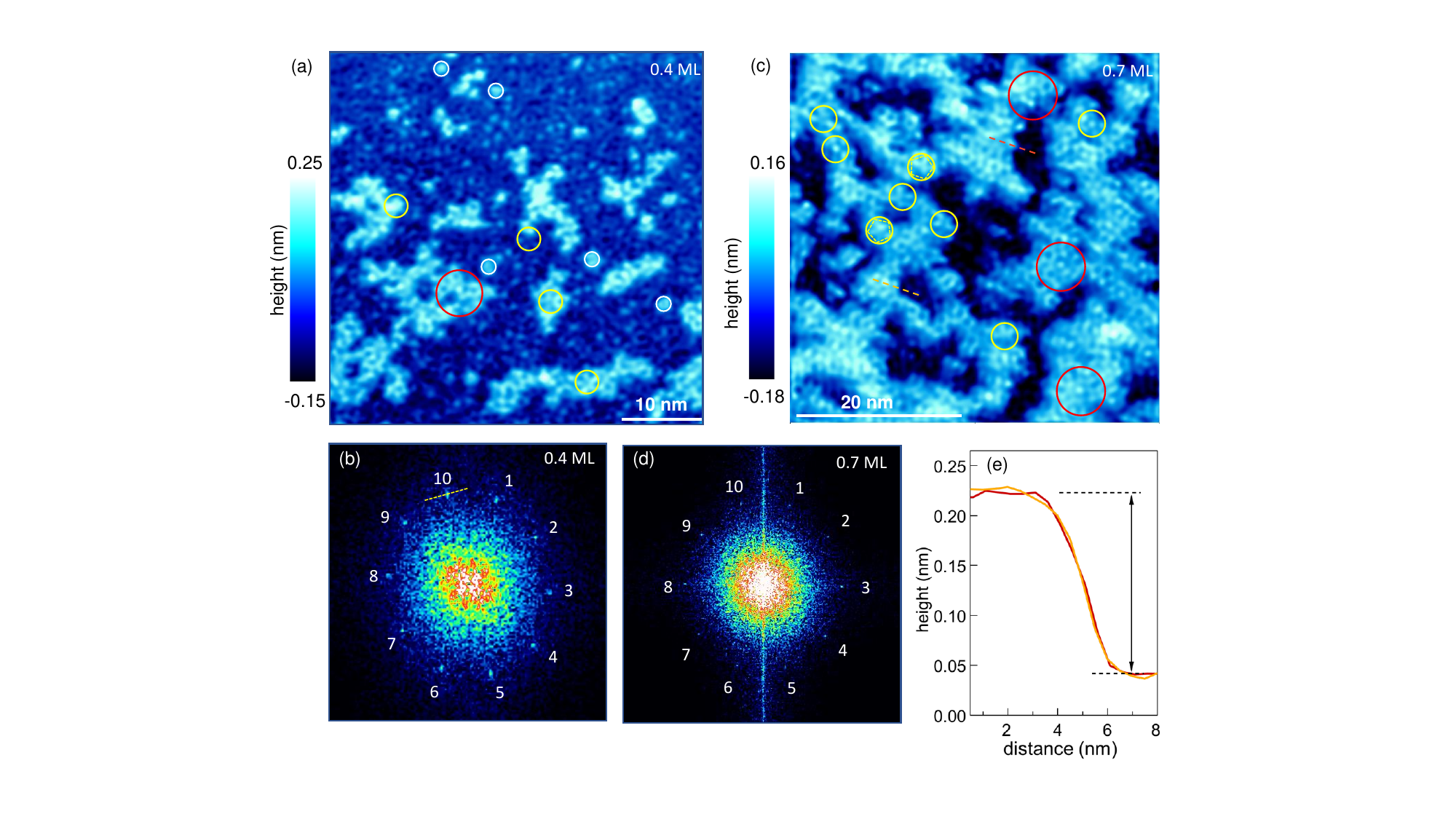}
 	\caption{(a) STM topography image of 0.4 ML Ga/\al\, ($I_T$ = 0.5 nA, $U_T$ = 1.0 V), where white, yellow, and red circles highlight the Ga white flower (GaWF), $\tau$-GaWF, and ring motifs, respectively.  (b) Fast Fourier transform (FFT) calculated for panel \textbf{a} by selecting only the Ga islands \srb{and masking  the substrate} using the height threshold procedure, the spots are numbered as 1-10. (c) STM topography image of 0.7 ML Ga ($I_T$= 0.7 nA, $U_T$ = 0.9 V), the motifs are highlighted as in panel \textbf{a}.	(d) FFT of the Ga islands in panel \textbf{c} \srb{after masking  the substrate} with the spots numbered. (e) The height profile along the  red and orange dashed lines in panel \textbf{c}.}
 	\label{submonostm}
 \end{figure*}

\subsection{\label{subsec:submonolayer_stm}Submonolayer Ga  on \al}
In Fig.~\ref{submonostm}(a), the deposition of Ga for $t_d$= 4 minutes results in the formation of condensed islands with a lateral size of $\sim$20 nm. The coverage turns out to be 0.4 ML.
~The height profiles show that their height  
~is similar to  the GaWFs, indicating that islands constitute  monolayer  Ga (the distribution is shown in  Fig.~S2(c)~\cite{supple}).  The formation of such islands -- in contrast to a dispersed phase reported for alkali metals on \al~\cite{Shukla2006} -- shows that the interaction between the adatoms is considerable. 

	\begin{figure*}[!t] 	
	\includegraphics[width=1.5\linewidth,keepaspectratio,trim={3cm 0.5cm 0 0 },clip]{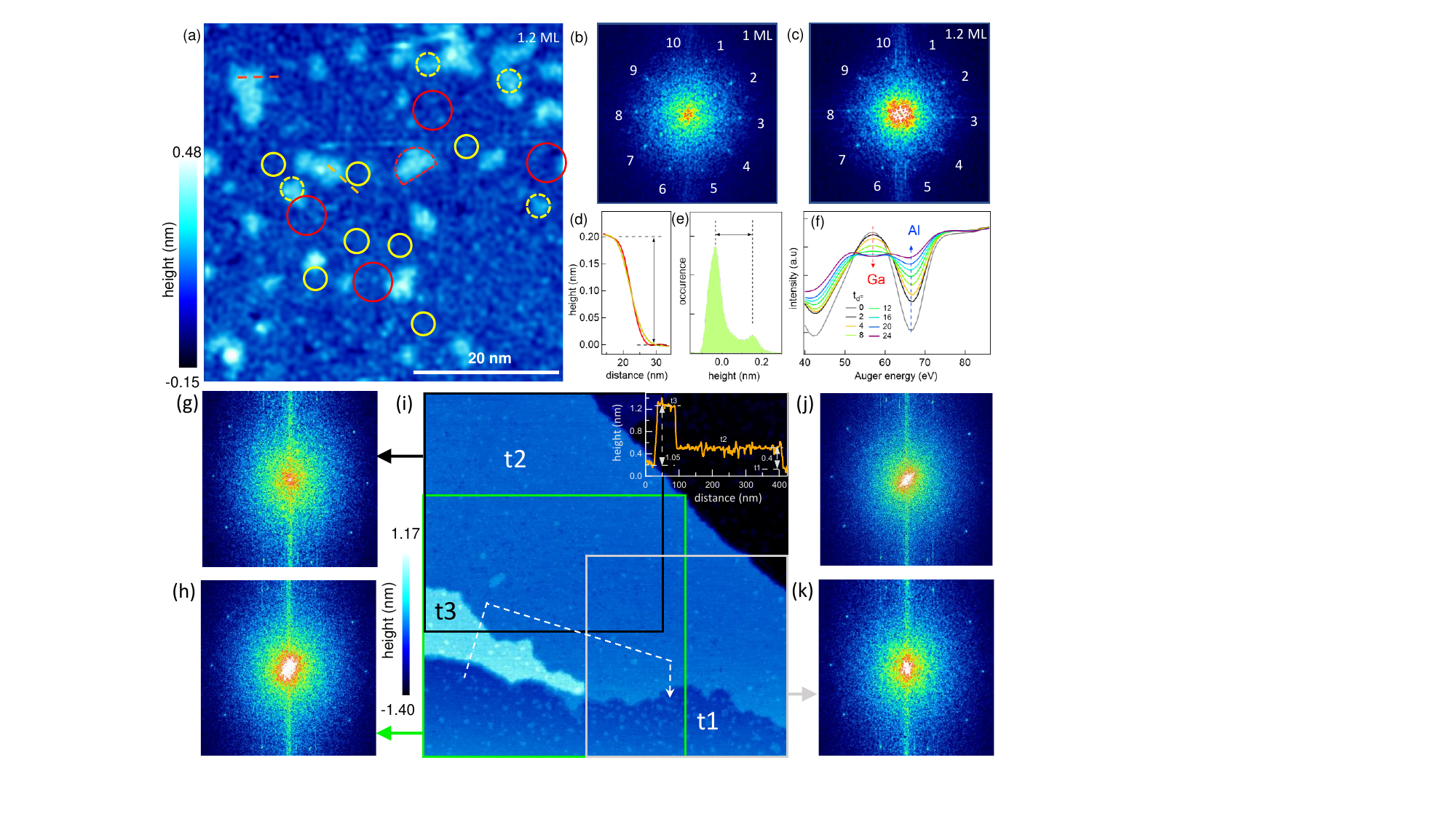}	
	\caption{(a) STM topography image of 1.2 ML Ga/\textit{i}-Al-Pd-Mn ($I_T$= 0.8 nA, $U_T$ = 0.5 V) showing formation of a uniform Ga monolayer.  The motifs of the monolayer are highlighted by circles of similar color, as shown in Fig.~\ref{submonostm}. The motifs in the bright condensed islands that represent Ga bilayer regions are shown by dashed circles.   FFT  of  (b) the monolayer region \srb{by masking  the bilayer regions} and (c)  the whole image in panel \textbf{a}.  (d) The height profile along the red and yellow dashed lines in panel \textbf{a}. (e) Height histogram of panel \textbf{a}. The  double-sided black arrows in panels \textbf{d} and \textbf{e} indicate the thickness of the second Ga layer. (f) Auger electron spectra as a function of deposition time ($t_d$) show an increase in the Ga \textit{MNN} signal (red arrow) and a decrease in the Al \textit{LMM} (blue arrow) signal.  (i) A large area (250 nm$\times$250 nm)  STM topography image of 1.2 ML Ga/\textit{i}-Al-Pd-Mn ($I_T$= 0.6 nA, $U_T$ = $-$1.9 V) and its (j) FFT. The inset of panel \textbf{i} shows the height profile along the white dashed line.   FFT of the different regions of  panel \textbf{i}, as indicated by the (g) black, (h) green, and (k) gray squares. }
	\label{monolayer}
\end{figure*}

 It is interesting to note here that besides the  GaWF motifs (white circles),  larger sized white flower motifs  [yellow circles in Fig.~\ref{submonostm}(a)] are observed. We refer to these as  $\tau$-inflated  GaWF or $\tau$-GaWF as their size scales with $\tau$ compared  to the GaWF motif.  
 ~In addition, red circles highlight a larger, more complicated motif with a bright center and an outer ring of pentagons.
 ~We refer to it as the ``ring'' motif.  In subsection ~\ref{subsec:motif_stm}, we elaborate on the origin of these motifs based on our DFT calculation.  
 
 To establish the quasiperiodic nature of  the Ga adlayer at 0.4 ML coverage, in Fig.~\ref{submonostm}(b) we show the FFT of the Ga island regions of the STM image \srb{by masking  the substrate using  the height threshold procedure}. The quasiperiodic nature of the islands is confirmed by the 10-fold FFT spots, with the angle subtended by two adjacent spots at the center being 36$^{\circ} \pm$ 2$^{\circ}$. In Fig.~S4(a) 
 ~of SM~\cite{supple}, the presence of all the spots is confirmed by the intensity profiles that show a peak that represents each spot. 

The STM topography image in Fig.~\ref{submonostm}(c) shows that for 0.7 ML  Ga coverage ($t_d$= 8 min), the condensed islands grow larger in size and merge together.  The $\tau$-inflated GaWF motifs (yellow circles) are more prevalent at this coverage. Additionally, their orientation is found to vary, 
~as indicated by the yellow dashed pentagons that are rotated by 36$^{\circ}$. 

The FFT in Fig.~\ref{submonostm}(d) [see also Fig.~S4(b)
~of SM~\cite{supple}] for  the Ga adlayer region  displays 10-fold spots with the angle subtended  at the center by the adjacent spots being 36$^{\circ}$$\pm$2$^{\circ}$. Thus, the FFT shows that the quasiperiodicity is retained up to this coverage.  The height profiles taken along the red and orange dashed lines shown in Fig.~\ref{submonostm}(c) are plotted in Fig.~\ref{submonostm}(e). Considering several such height profiles, the average height turns out to be 0.18$\pm$0.03 nm, in agreement with that determined for 0.1-0.4 ML. Note that the thickness of the Ga monolayer is somewhat smaller compared to other quasiperiodic layers of elemental metals on \al, such as Sn~(0.2 nm)~\cite{Singh2020} and Pb~(0.23 nm)~\cite{Ledieu2008,Ledieu2009}. A possible reason is the smaller size of the Ga atom (empirical atomic diameter being 0.26 nm~\cite{Slater1964}) compared to Sn (0.29 nm) and Pb (0.36 nm).

\subsection{Quasiperiodicity of monolayer 
	gallium}
\label{subsec:monolayer_stm}
Figure~\ref{monolayer}(a) illustrates the STM topography image for $t_d$= 16 minutes, in which the substrate is uniformly covered by the Ga monolayer. Note that the characteristic quasiperiodic motifs such as $\tau$-GaWF and ring -- similar to those observed in Fig.~\ref{submonostm} --  are also observed on  the Ga monolayer, these are highlighted by similar colored circles  in Fig.~\ref{monolayer}(a).   

 Curiously, in this figure, isolated bright  condensed islands are observed.  In Fig.~\ref{monolayer}(d), the height profiles along the red and yellow dashed lines in Fig.~\ref{monolayer}(a)  show that the average   height of these islands  relative to the monolayer is  0.19$\pm$0.02 nm. This is nearly similar to the monolayer's height discussed in the previous section, and thus  the bright islands can be related to Ga bilayer regions. This is further supported by the height histogram in Fig.~\ref{monolayer}(e), which displays two peaks corresponding to the mono- and bilayer, the separation between which yields a similar height of the latter (horizontal double-sided arrow).  From the area under the two peaks of the height histogram, we find that the bilayer forms over an  area  that is $\approx$20\% of the total area. Thus, the coverage for this deposition turns out to be 1.2 ML.  
  
Fig.~\ref{monolayer}(b) displays the FFT of only the monolayer (i.e., excluding the bilayer regions), while Fig.~\ref{monolayer}(c) displays the FFT of the entire image. 10-fold spots are observed in both. Notably, from a  comparison of their intensity profiles in  Fig.~S5 of SM~\cite{supple},
 ~it is apparent that the intensity 
 ~of the spots is larger with the inclusion of the bilayer regions,  indicating  an improvement in the quality of the FFT.  \srb{In addition,   close scrutiny reveals the characteristic quasiperiodic motifs such as $\tau$-GaWF and an incomplete ring on the bilayer islands  [dashed circles in Fig.~\ref{monolayer}(a)], 
~indicating the possible existence of  quasiperiodic structural correlations in some regions of the bilayer. }

In Fig.~\ref{monolayer}(f), the Ga MNN Auger electron spectroscopy signal at 55 eV  increases with a concomitant decrease of the substrate Al LMM signal at 68 eV as a function of $t_d$.  This shows that the Ga atoms are deposited incrementally on the \al\, surface. No change in the position of either of the peaks is observed over the whole $t_d$ range (see the  vertical dashed arrows), indicating the absence of any surface alloying or chemical bonding  with the substrate. 

\srb{A STM topography image in Fig.~\ref{monolayer}(i) for 1.2 monolayer coverage  ($t_d$=  16 min) that spans a larger length scale (250nm$\times$250 nm) 
	shows  three adjacent terraces of the substrate.  The bottom terrace is marked as t1 (blue), while the other terraces that exhibit  step heights of  0.4$\pm$0.02 nm (= $L$) and   1.05$\pm$0.02 nm (= $2L+S$, 
	~$L$ and $S$ are the basic heights related by $L= \tau\times S$)  are referred to as t2 (light blue ) and t3 (whitish blue), respectively; see the inset for the height profile along the white dashed line.  $L$ and  $(2L+S)$ are the characteristic step heights of the  substrate~\cite{Sharma2007,Schaub1994,
		Shen1999}; their presence indicates  that these terraces are uniformly covered by the Ga monolayer (the small bright isolated islands are the bilayer regions).    The quasiperiodicity of Ga on  these terraces is established by the presence of sharp 10-fold spots in the FFT, as shown in 	Fig.~\ref{monolayer}(g) for an overlapping area of t2 and t3 terraces (black square). In  Fig.~\ref{monolayer}(h), FFT is shown for an overlapping area of all the three  terraces (green square), while  Fig.~\ref{monolayer}(k)  shows it for  t1 and t2 (gray square).  The whole image also shows  FFT spots [Fig.~\ref{monolayer}(j)]. 
	~Thus, the Ga monolayer is  quasiperiodic over  t1, t2, and t3 terraces, whose combined area is  0.0525~$\mu$m$^2$.  The largest length scale over which quasiperiodicity is observed here is $\sim$350 nm, which is the diagonal of the image in Fig.~\ref{monolayer}(i). The signature of  long range quasiperiodicity of the Ga monolayer is thus obtained over a range larger than the previous reports in the literature for other elements~\cite{Sharma2013, Ledieu2008,  Singh2020, Singh2023, Smerdon2008, Shimoda2004}.}

	\begin{figure*}[!t] 
	\includegraphics[width=1.2\linewidth,keepaspectratio,trim={6cm 0 4cm 0cm },clip]{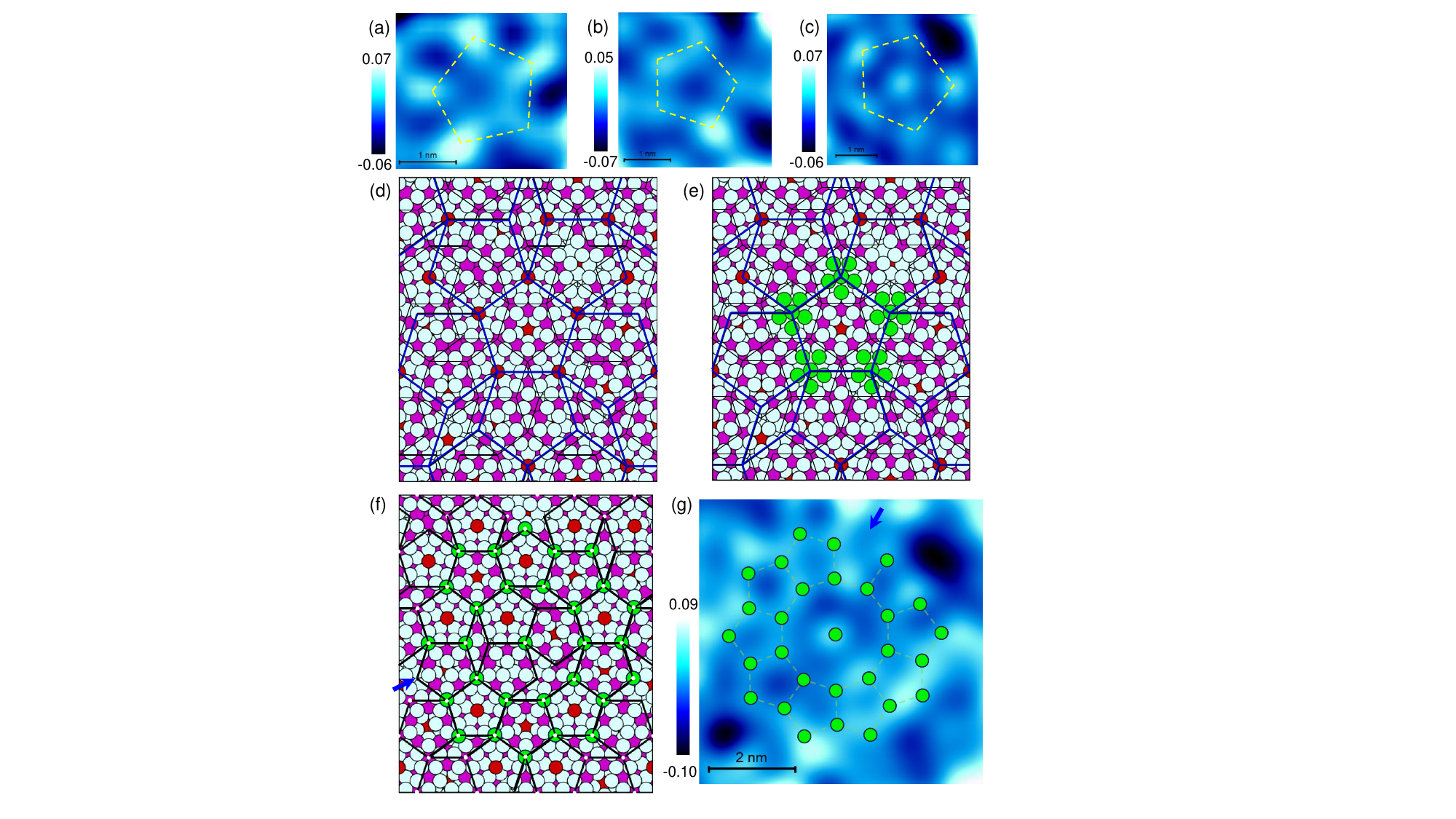}	
	\caption{Zoomed STM images  of the $\tau$-GaWF motif with a dark center  observed for  the Ga (a) monolayer and (b)  bilayer. (c)  A $\tau$-GaWF  motif with a bright center. The color scale representing the 	height in nm is shown on the left side of panels \textbf{a-c} and \textbf{f}, zero corresponds to the bearing height. (d) The atomic structure of the surface plane   of  5-fold \al\, derived from the 5/3 approximant (5.32 nm$\times$ 6.25 nm), \srb{the thick  blue (black) lines show the $\tau$-P1 (P1) tiling.} The white, magenta, and red colored circles represent the Al, Pd, and Mn atoms, respectively. (e) Same as panel \textbf{d} with DFT based  30-Ga atom (green circles)  model  representing the $\tau$-GaWF motif on the \al\, surface. \srb{(f) A  model for the ring motif, the white dots show all possible adsorption sites on the Pd atoms at the center of the 5-Al cluster. The blue arrow shows a vertex of the outer pentagon that is not a favorable site. 	The black lines show the P1 tiling that is shifted compared to panel \textbf{d}. (g) The green circles show the adsorbed Ga atoms of the model connected by green dashed lines 
			overlaid on the zoomed ring motif from STM. Both panels \textbf{f} and \textbf{g} have same length scale. 
}}
	\label{motif} 
\end{figure*} 

\subsection{\label{subsec:motif_stm} DFT based models for the $\tau$-GaWF and ring motifs }
The STM images in Figs.~\ref{submonostm}-\ref{monolayer} show that the  $\tau$-GaWF motif with a dark center is most frequently observed, 
whose zoomed views are  shown in  Figs.~\ref{motif}(a,b). Some of these motifs are also observed with a bright center [Figs.~\ref{motif}(c)].  To show that these motifs are  $\tau$ inflated compared to the GaWF motifs, we determine the edge length  of the pentagon  shown by  yellow dashed lines in Figs.~\ref{motif}(a-c). Its average value is  1.26$\pm$0.1 nm, the corresponding distribution is shown in Fig.~S6
~of SM~\cite{supple}.  The experimental ratio of the side length of $\tau$-GaWF and the GaWF motifs is $\frac{1.26}{0.79}$= 1.59, which is close to $\tau$ (= 1.618). Thus, STM shows that  the $\tau$-GaWF motif is indeed   $\tau$ inflated with respect to the GaWF motif. 

\srb{The size of this motif can be approximately estimated  by the diameter of a circle circumscribing it drawn such that the $\tau$-GaWF motif is  contained in it. The distribution of the diameter, considering several such motifs, is plotted as a histogram in Fig.~S7(a)~\cite{supple}. The average diameter of the $\tau$-GaWF motif turns out to be  3.2$\pm$0.2 nm. This is approximately $\tau$ scaled with respect to the diameter of the GaWF (2$\pm$0.1 nm), thus justifying the name ``$\tau$-GaWF". }

Fig.~\ref{motif}(d) shows the atomic structure of the surface plane of the 5-fold \al\, derived from the 5/3 approximant. To model larger sized motifs by DFT,  the $\tau$-P1 tiling is more appropriate (blue lines),  where the size of the tiles is $\tau$ times larger with the edge length  $a_1$= 1.255 nm (=$\tau$$\times$$a_0$, $a_0$= 0.776 nm) \srb{compared to the P1 tiling (black lines)}.  Here, all the surface Mn atoms are at the vertices of the $\tau$-P1 tiling, but not all vertices of the $\tau$-P1 tiling are occupied by Mn atoms.  In the central P1 pentagon, all 5 vertices are occupied by Mn atoms.  At the center of this $\tau$-P1 pentagon, a dip can  be observed. In previous studies, such a surface dip was also characterized also by a charge density minimum (see e.g., Fig.~3 in Ref.~\onlinecite{Kraj2010}).

 If around the Mn atoms of the $\tau$-P1 pentagon the 6-Ga  clusters grow,  a new Ga superstructure  can be observed that would represent $\tau$-GaWF, as shown in Fig.~\ref{motif}(e). This model is supported by the excellent agreement of  $a_1$ estimated from the side length of the $\tau$-GaWF motifs from STM  (1.26$\pm$0.1 nm)  with the theoretical value (1.255 nm). Also, this model accurately reflects the distinct petals  in  the $\tau$-GaWF motif observed from STM.  On \al\,  surfaces, the dips could be filled by Ga adatoms,  and the dark center of the $\tau$-GaWF could disappear. This could explain the  bright centered $\tau$-GaWF motif  in Figs.~\ref{submonostm}(c). On the \al\, surface, $\tau$-P1 pentagons with Mn atoms at all 5 vertices exist in varying orientations. This explains why $\tau$-GaWF clusters could be rotated relative to one another, as seen in Figs.~\ref{motif}(a,b) and Fig.~\ref{submonostm}(c).

\begin{figure*}[thb] 
	\centering
	\includegraphics[width=\linewidth, keepaspectratio,trim={0cm 7cm 2cm 3cm},clip]{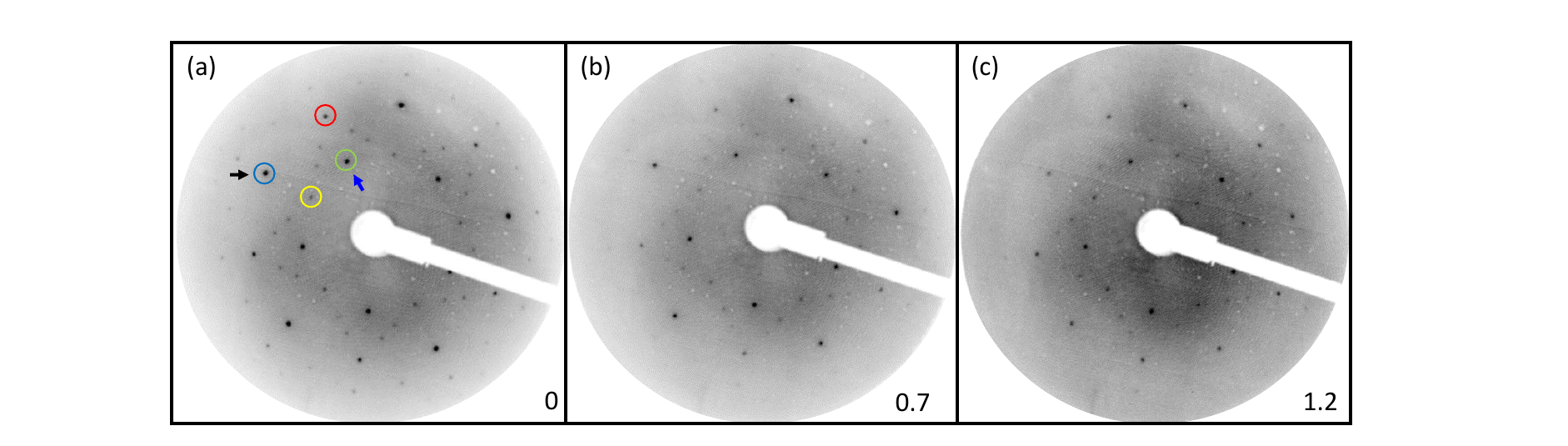} 
	\caption{LEED patterns at a beam energy ($E_p$) of 81 eV for (a) 0,  (b) 0.7,   and (c) 1.2 ML 
		~Ga/\textit{i}-Al-Pd-Mn in an inverted gray scale, the coverage is indicated  at the bottom right corner of each panel.} 
\label{leedsimilar}
\end{figure*}

A zoomed view of the ring  motif  (red circles  in  Figs.~\ref{submonostm}- \ref{monolayer}), which comprises of a nearly circular arrangement of  outer pentagons,  is shown in Fig.~\ref{motif}(g).  The center of the ring motif is bright and is surrounded by a relatively darker region. \srb{An estimate of the average diameter of this motif  is 5.2$\pm$0.4 nm, indicating that it is approximately $\tau$ inflated with respect to the $\tau$-GaWF motif (3.2 nm).  The distribution of the diameters is shown in  Fig.~S7~\cite{supple}.} 

	\srb{In Fig.~\ref{motif}(f), we present a DFT based model of the ring motif  that consists of an outer  ring of  pentagons  formed by Ga atoms  that occupy the favorable Pd sites at the middle of the 5-Al cluster that occur at the vertices of the pentagons of the P1 tiling shown by black lines.  All possible favorable sites for Ga adsorption are shown by white dots in Fig.~\ref{motif}(f), while the green circles in Fig.~S3(c) of SM represent a complete ring motif.  The bright center of the ring motif also arises from a Ga adatom adsorbed at the center Pd atom surrounded by the 5-Al cluster. This is the common vertex of three inner P1 pentagons, however two vertices of each of these pentagons are not energetically favorable.  To be noted is that 6-Ga clusters on the Mn atoms at the center of the alternate pentagons could form a GaWF, but it is somewhat energetically less favored compared to only the outer 5-Ga considered in the model (Table~\ref{table1}).  A satisfactory  agreement can be noticed in both the atomic positions and motif size when this model is superimposed on the ring motif depicted in Fig.~\ref{motif}(g) subsequent to a 120$^{\circ}$ rotation.  It is worth noting that the majority of the vertices comprising the outer ring of 10 pentagons are energetically favorable (as indicated by white dots). However, there is one unfavorable vertex, denoted by a blue arrow in Figs.~\ref{motif}(f,g), which lacks both the Pd atom and the 5-Al cluster.  Additionally, a neighboring vertex of the ring motif in Fig.~\ref{motif}(g) is unoccupied for stochastic reasons.} 

	\subsection{\label{subsec:submonolayer_LEED}Low energy electron diffraction study of Ga/\textit{i}-Al-Pd-Mn}
	
		Figure~\ref{leedsimilar}(a) displays the LEED pattern of  \al\, that has different sets of sharp 5-fold spots  in agreement with the literature~\cite{Gierer1997, Shukla2009,  Singh2020}. The most intense inner and outer sets of spots are indicated by blue and black arrows, respectively. \srb{These spots with similar relative intensities between the 5-fold sets  are  clearly visible  for 1.2  ML coverage in Fig.~\ref{leedsimilar}(c)], where the surface is uniformly covered by Ga, as shown by STM in Fig.~\ref{monolayer}, also see Fig.~S8 for other $E_p$ values. The spots are also visible for the submonolayer coverage of 0.7 ML [Figs.~\ref{leedsimilar}(b)]. The intensity profiles of the inner and outer sets of spots for 1.2 ML LEED pattern in Fig.~S9(a) 
		~of SM~\cite{supple}  show that the angular position (separation between spots being 36$^{\circ}$$\pm$1$^{\circ}$) and the width of the spots remain nearly unchanged with respect to the substrate. The profiles along three radial directions in Figs.~S9(b-d) show similar $k_{\parallel}$ for both sets of spots, indicating  an unchanged quasi-lattice parameter of the Ga adlayer compared to the substrate.} There is no evidence of splitting or broadening of the spots indicating the absence of surface defects or steps of multiple heights~\cite{VanHove1979}. 
				It may be noted that rotational epitaxy instead of quasiperiodicity has been reported for  several metals deposited on \al\, such as  Al~\cite{Bolliger2001}, Ag~\cite{Fourne2003}, Fe~\cite{Weisskopf2005}, and Ni~\cite{Weisskopf2006}. In these cases, the adlayer manifests as five crystalline domains that mirror the symmetry of the substrate, and these domains add extra spots to the LEED pattern. Such extra spots are absent for the Ga monolayer  for the  whole $E_p$ range, as shown by  a series of patterns from 30 eV $<$$E_p$$<$ 180 eV at a step of 2 eV in video files named ``Ga0.7ML" and ``Ga1.2ML" for  0.7 and 1.2 ML,   respectively of the SM~\cite{supple}.  This rules out the possibility of the formation of Ga  domains  with a particular orientational relationship with the substrate. \srb{Thus, our electron diffraction study shows that the Ga monolayer exhibits quasiperiodic order within the coherence length scale of the LEED optics, which is 10-20 nm~\cite{Hattori2003}.} There is however a decrease in the  intensity of  the Ga adlayer spots,  indicating an increase in disorder and roughness.  This statement is supported by the fact that the roughness of the \srb{monolayer (0.05 nm), as measured by STM, is  larger compared to the substrate (0.02 nm).} 
		\begin{figure}[tb] 
	\centering
	\includegraphics[width=0.6\textwidth,keepaspectratio,trim={3cm -1cm 11cm 0 },clip]{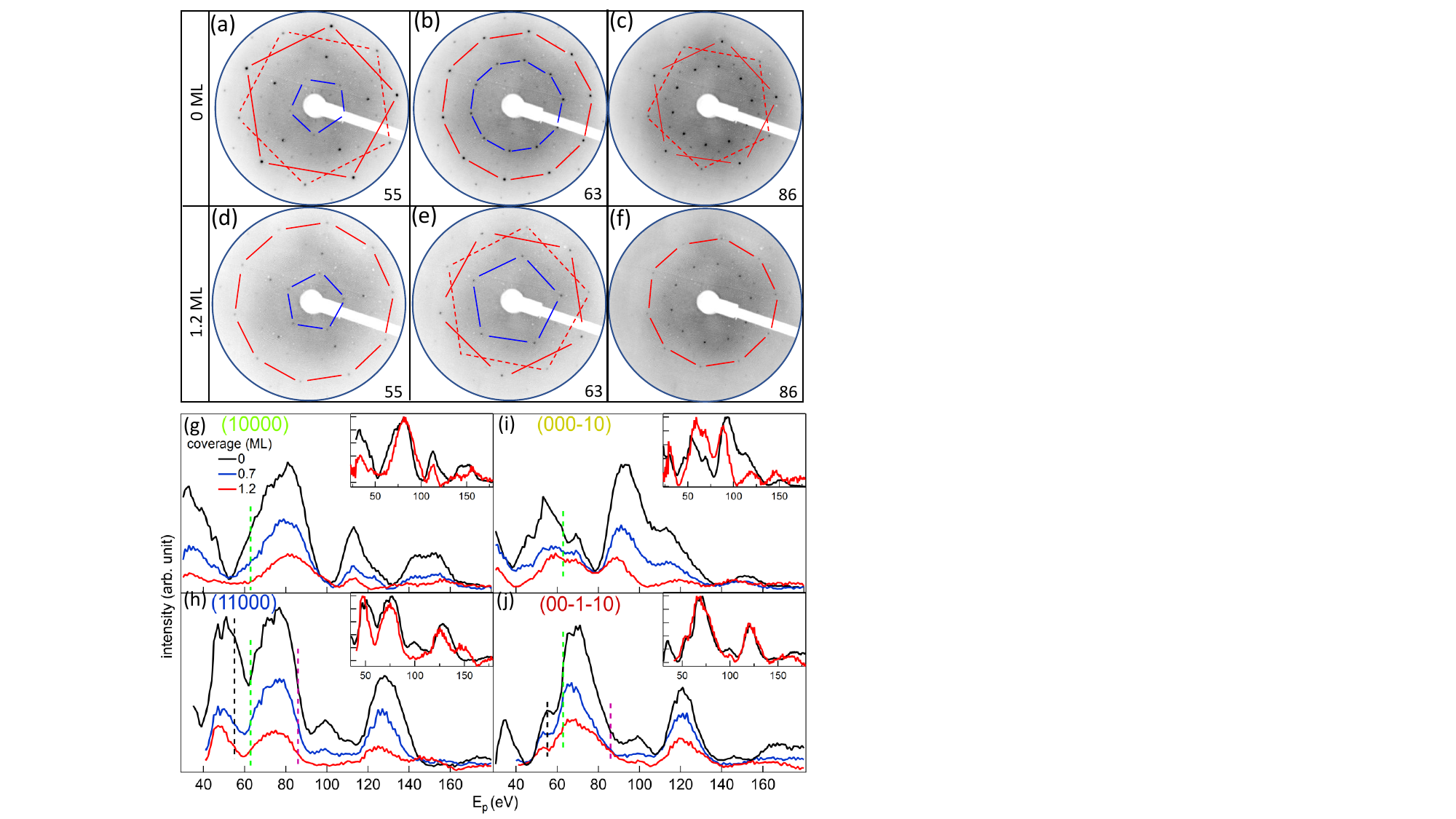}		
	\caption{LEED patterns of (a-c) \textit{i}-Al-Pd-Mn, and (d-f) 1.2 ML Ga/\textit{i}-Al-Pd-Mn for three different \textit{E$_p$} values are shown at the bottom right corner in eV. The differences are highlighted by blue and red lines.  Experimental \textit{I-V} curves for 0.7  and 1.2 ML  Ga/\al\, compared with  \textit{i}-Al-Pd-Mn (0 ML) for different spots (g) (10000), (h) (11000), (i) (000$\overline{1}$0), and (j) (00$\overline{1}$$\overline{1}$0). These spots are marked in Fig.~\ref{leedsimilar}(a)  by green, blue, yellow, and red circles, respectively. The insets compare the 1.2 ML \textit{I-V} with the substrate after normalizing the most intense peak to the same height.}
	\label{leediv}
\end{figure}	
	
\srb{Although the LEED patterns of the Ga adlayer resemble the substrate  in Fig.~\ref{leedsimilar}, for some $E_p$, the relative intensities of the two sets of 5 fold spots are different [Figs.~\ref{leediv}(a-f)]. For example, at $E_p$= 55 eV in Fig.~\ref{leediv}(d), the outer ring  has nearly equal intensities of the two sets (thus appearing like a decagon, red lines), while for the substrate in Fig.~\ref{leediv}(a) one set is more intense compared to the other (thus appearing as two 36$^{\circ}$ rotated pentagons, shown by solid and dashed red lines). This is also observed at 86 eV. Moreover, at 55 eV, the inner pentagon highlighted by blue lines is 36$^{\circ}$ rotated between the two. At 63 eV, the inner (outer) set of spots of the Ga layer portrays a blue (solid and dashed red)  pentagon, while for the substrate,  the spots are of equal intensity.}

\srb{To understand these differences, the  intensity ($I$) of a LEED spot as a function of $E_p$ -- called \textit{I-V} curve~\cite{VanHove1979,Heinz1995}  -- has been shown in Figs.~\ref{leediv}(g-j).  The \textit{I-V} curves of the substrate for different spots, such as  (10000),  (11000),  (000$\overline{1}$0), and (00$\overline{1}$$\overline{1}$0) are in agreement with the earlier reports~\cite{Heinzig2002, Singh2020}. Although a comparison with the substrate shows  a fair qualitative closeness in the positions of the main peaks,  there are perceptible differences in particular for  (000$\overline{1}$0) and (11000) spots, which are apparent  in the insets where  their most intense peak has been normalized to the same height. The difference  in the shapes of the  \textit{I-V} curves could arise from  the structural difference caused by preferred adsorption positions of Ga on \al\ as shown by DFT  as well as non-structural parameters such as change in the surface potential~\cite{VanHove1979}.  The differences  in the LEED patterns between the Ga adlayer and the substrate in Figs.~\ref{leediv}(a-f) can be explained by the difference in the shapes of the \textit{I-V} curves [dashed lines in Figs.~\ref{leediv}(g-j)].}  \\

\section{Conclusions}
	In conclusion, our present study utilizing scanning tunneling microscopy (STM), low energy electron diffraction (LEED), and density functional theory (DFT) demonstrates  that Ga monolayer on  \al\, exhibits   quasiperiodic order  at room temperature.  The FFTs of the STM images exhibit the characteristic quasiperiodic spots, while the real space images reveal motifs such as Ga white flower (GaWF) and $\tau$-inflated GaWF ($\tau$-GaWF), where $\tau$ is the golden mean. A larger sized ring  motif made up of  pentagons in a circular congregation with a bright center is also observed. STM shows  that the quasiperiodic order persists over a length scale of  $\sim$350 nm. The quasiperiodic nature of the Ga monolayer is further supported by the 5-fold LEED patterns observed  across the entire range of  the beam energy.  Auger electron spectroscopy  provides additional evidence that the  Ga adlayer does not exhibit any  surface alloying or chemical reactions with the substrate. 
	
	Our DFT calculations indicate that Ga prefers to adsorb on \al\ at two specific sites: (i) the vertices of the Penrose P1 tile located at the center of a cluster consisting of 5 Al atoms that has a Pd atom at the center but positioned below the surface plane, and (ii) the Mn atom positioned at the center of the P1 tile.   The  GaWF motif is represented by a pentagonal outer ring comprising 5 Ga atoms adsorbed at the center of 5-Al clusters and a pentagonal inner cluster consisting of 6 Ga atoms (6-Ga) surrounding the central Mn atom.  The 6-Ga clusters  arranged on the $\tau$-P1 tiling serve as a model for the $\tau$-GaWF motif,  \srb{while the ring motif is represented by Ga atoms that adsorb at the reactive centers of 5-Al clusters above the marked Pd sites. Excellent agreement is observed between the side length and the diameter of the motifs between the experiment and the  DFT-based models.} 	Our research reveals a previously unknown characteristic of gallium, i.e., its quasicrystallinity. It is intriguing that Ga exhibits quasiperiodic order despite the substrate temperature being near  its melting point; this characteristic sets it apart from other systems.
		
\section{Acknowledgments}
M.K. is thankful for the support from the Slovak Grant Agency VEGA (No. 2/0144/21) and APVV (No. 19-0369). We are grateful to M. Mihalkovi\v{c} and K. Pussi for  insightful discussions. We thank D. L. Schlagel and T. A. Lograsso for providing us with the \al\, substrate.  Special appreciation is extended to V. K. Singh for his suggestions regarding the STM measurement. 
	
	 \noindent e-mail addresses:\\$^{*}$marian.krajci@savba.sk, \\$^{\dagger}$barmansr@gmail.com 
		%
		
	\end{document}